\documentclass{aa}  

\usepackage{graphicx}
\usepackage{txfonts}
\usepackage{longtable} 
\begin{document}

   \title{The role of binaries in the enrichment of the early Galactic halo.}

   \subtitle{I. $r$-process-enhanced metal-poor stars}

   \author{T. T. Hansen
          \inst{1}
          \and
          J. Andersen\inst{2,3}\and B. Nordstr{\"o}m\inst{2,3}\and
          T. C. Beers\inst{4} \and J. Yoon\inst{4} \and L. A. Buchhave\inst{5,6} }

   \institute{Landessternwarte, ZAH, Heidelberg University,
              K{\"o}nigstuhl 12, Heidelberg, D-69117, \\
              \email{thansen@lsw.uni-heidelberg.de}
   \and
             Dark Cosmology Centre, The Niels Bohr Institute, Copenhagen University,\\
             Juliane Maries Vej 30, DK-2100 Copenhagen, Denmark\\
             \email{ja@nbi.ku.dk, birgitta@nbi.ku.dk}
   \and
    Stellar Astrophysics Centre, Department of Physics and Astronomy,\\
    Aarhus University, Ny Munkegade 120, DK-8000 Aarhus C, Denmark
\and
Department of Physics and JINA Center for the Evolution of the Elements, \\ University of Notre Dame, Notre Dame, IN, 46556, USA\\
\email{tbeers@nd.edu,jyoon4@nd.edu}
\and
Harvard-Smithsonian Center for Astrophysics, Cambridge, MA 02138, USA
\and
Centre for Star and Planet Formation, Natural History Museum of Denmark,
University of Copenhagen, DK-1350 Copenhagen, Denmark\\
\email{buchhave@astro.ku.dk}           
             }

   \date{}

 
  \abstract
   {The detailed chemical composition of most metal-poor halo
     stars has been found to be highly uniform, but a minority of stars
     exhibit dramatic enhancements in their abundances of heavy
     neutron-capture elements and/or of carbon. The key question for Galactic
     chemical evolution models is whether these peculiarities reflect the
     composition of the natal clouds, or if they are due to later (post-birth)
     mass transfer of chemically processed material from a binary
     companion. If the former case applies, the observed excess of certain
     elements was implanted within selected clouds in the early ISM from a
     production site at interstellar distances.} 
  {Our aim is to determine the frequency and orbital properties of binaries
    among these chemically peculiar stars. This information provides the basis
    for deciding whether local mass transfer from a binary companion is
    necessary and sufficient to explain their unusual compositions. This paper
    discusses our study of a sample of 17 moderately ($r$-I) and highly
    ($r$-II) $r$-process-element enhanced VMP and EMP stars.}  
  {High-resolution, low signal-to-noise spectra of the stars were obtained at roughly
    monthly intervals over eight years with the FIES spectrograph at the
    Nordic Optical Telescope. From these spectra, radial velocities with an
    accuracy of $\sim$100 m~s$^{-1}$ were determined by cross-correlation
    against an optimized template.} 
  {Fourteen of the programme stars exhibit no significant radial-velocity
    variation over this temporal window, while three are binaries with orbits of
    typical eccentricity for their periods, resulting in a normal binary
    frequency of $\sim18\pm$6\% for the sample.} 
  {Our results confirm our preliminary conclusion from 2011, based
    on partial data, that the chemical peculiarity of the $r$-I and $r$-II
    stars is not caused by any putative binary companions. Instead, it was
    imprinted on the natal molecular clouds of these stars by an external,
    distant source. Models of the ISM in early galaxies should account for
    such mechanisms.}  
   \maketitle
%

\section{Introduction}

\begin{table*}
\caption{$r$-I and $r$-II stars monitored for radial-velocity variation}
\label{tbl-1}
\centering
\begin{tabular}{lrrrrrrl}
\hline\hline
Stellar ID & RA (J2000) & Dec (J2000) & $V$ & $B-V$ & $\mathrm{[Fe/H]}$ &
$\mathrm{[Eu/Fe]}$ & Class \\
\hline
\object{HD~20}         &00:05:15 &$-$27:16:18 & 9.24 &0.54 &$-$1.58 &$+$0.80 & $r$-I  \\
\object{CS~29497$-$004}&00:28:07 &$-$26:03:03 &14.03 &0.70 &$-$2.81 &$+$1.62 & $r$-II \\
\object{CS~31082$-$001}&01:29:31 &$-$16:00:48 &11.67 &0.77 &$-$2.78 &$+$1.66 & $r$-II \\
\object{HE~0432$-$0923}&04:34:26 &$-$09:16:50 &15.17 &0.73 &$-$3.19 &$+$1.25 & $r$-II \\
\object{HE~0442$-$1234}&04:44:52 &$-$12:28:46 &12.91 &1.08 &$-$2.41 &$+$0.52 & $r$-I  \\
\object{HE~0524$-$2055}&05:27:04 &$-$20:52:42 &14.01 &0.88 &$-$2.58 &$+$0.49 & $r$-I  \\
\object{HE~1044$-$2509}&10:47:16 &$-$25:25:17 &14.35 &0.67 &$-$2.89 &$+$0.94 & $r$-I  \\
\object{HE~1105$+$0027}&11:07:49 &$+$00:11:38 &15.65 &0.39 &$-$2.42 &$+$1.81 & $r$-II \\
\object{HE~1127$-$1143}&11:29:51 &$-$12:00:13 &15.89 &0.68 &$-$2.73 &$+$1.08 & $r$-II \\
\object{HE~1219$-$0312}&12:21:34 &$-$03:28:40 &15.94 &0.64 &$-$2.81 &$+$1.41 & $r$-II \\
\object{HE~1430$+$0053}&14:33:17 &$+$00:40:49 &13.69 &0.59 &$-$3.03 &$+$0.72 & $r$-I  \\
\object{HE~1523$-$0901}&15:26:01 &$-$09:11:38 &11.13 &1.06 &$-$2.95 &$+$1.82 & $r$-II \\
\object{CS~22892$-$052}&22:17:01 &$-$16:39:26 &13.21 &0.80 &$-$2.95 &$+$1.54 & $r$-II \\
\object{HE~2224$+$0143}&22:27:23 &$+$01:58:33 &13.68 &0.71 &$-$2.58 &$+$1.05 & $r$-II \\
\object{HE~2244$-$1503}&22:47:26 &$-$14:47:30 &15.35 &0.66 &$-$2.88 &$+$0.95 & $r$-I  \\
\object{HD~221170}     &23:29:29 &$+$30:25:57 & 7.71 &1.02 &$-$2.14 &$+$0.85 & $r$-I  \\
\object{CS~30315$-$029}&23:34:27 &$-$26:42:19 &13.66 &0.92 &$-$3.33 &$+$0.72 & $r$-I  \\
\hline
\end{tabular}
\tablebib{$V$ and $B-V$ are from \citet{beers2007}, except for
HD~221170, which is taken from \citet{christlieb2004}, and
HE~1127$-$1143, which is from \citet{henden2015}.
$\mathrm{[Fe/H]}$ and $\mathrm{[Eu/Fe]}$ are from \citet{barklem2005},
except for HE~1523$-$0901, where values from \citet{frebel2007} are listed.}
\end{table*}

The pioneering HK survey of Beers, Preston, \& Shectman \citep{beers1985,
beers1992} demonstrated the existence of substantial numbers of stars
with metallicities well below those of globular clusters, and opened a
new observational window on the epoch of early star formation in the
Galactic halo. Modern surveys of metal-poor stars has greatly expanded
the numbers of such stars known at present \citep[see][for a
recent review]{ivezic2012}, and spectroscopy with 8-metre class telescopes have
demonstrated a surprisingly uniform abundance pattern in the majority of
them \citep{cayrel2004,bonifacio2009,aoki2013,yong2013,roederer2014a}. 

Since very metal-poor (VMP; $\mathrm{[Fe/H]} < -2.0$) and extremely
metal-poor (EMP; $\mathrm{[Fe/H]} < -3.0$) halo stars probe the earliest
epochs of chemical evolution in the Galaxy, their elemental-abundance
patterns reflect the products of the primary heavy-element synthesis and
enrichment processes in the early Galaxy. However, significant samples
of chemically peculiar stars have been identified, in particular among
the VMP and EMP stars. Because anomalies affecting a single element or
group of elements must be produced by a small number of nucleosynthesis
events (including possible single events), the chemically peculiar stars
provide the opportunity to characterize the progenitors and enrichment
processes that produced the abundance patterns in the long-lived
metal-poor stars seen today.

However, in the relatively more metal-rich Population I and II
stars, some chemical anomalies are known to be due to evolution in a
close binary system (e.g. Ba and CH stars), in which a higher-mass
binary companion has evolved to the asymptotic giant-branch (AGB) stage.
Through processes involving Roche-lobe overflow and/or wind accretion,
the higher-mass star is expected to have contaminated the envelope of
its surviving companion with any elements that formed during its final
evolutionary stages. This mechanism could explain the presence of
elements produced in the early Universe by stars more massive than
$\sim$0.7 M$_{\sun}$ (i.e. with main-sequence lifetimes less than the
Hubble time), but does of course require the presence of a binary
companion, which by now is presumably a white dwarf or neutron star.   

Hence, for the chemically peculiar VMP and EMP stars, a crucial first
step is to establish whether any such putative binary companions
actually exist, and to determine their main orbital parameters (i.e.
period, semi-major axis, and eccentricity). This requires
radial-velocity monitoring of adequate precision and duration for a
sufficiently large sample of stars. High spectroscopic resolution ($R
~\ga30,000$) allows one to determine radial velocities efficiently by
cross-correlating even a relatively low signal-to-noise ratio
(S/N) spectrum with an appropriate template, while long-term thermal
and mechanical stability of the spectrograph employed is needed in
order to detect variations with amplitudes below 1 km~s$^{-1}$. As we
demonstrate below, the FIES spectrograph at the 2.56-m Nordic Optical
Telescope (NOT), which we have used to carry out our programme, meets
these requirements. 

This paper is the first in a series that explores the nature of three
broad classes of chemically peculiar metal-poor halo stars. Here we
consider the nature of VMP and EMP stars that exhibit moderate-to-large
enhancements of elements associated with the rapid neutron-capture
process, the so-called $r$-I and $r$-II stars, respectively. Other
papers in this series consider samples of carbon-enhanced metal-poor
(CEMP) stars \citep[see][]{beerschristlieb2005}: the CEMP-no stars
(which exhibit no over-abundances of neutron-capture process elements;
Paper~II), and the CEMP-$s$ stars (which exhibit over-abundances of
elements originating in the slow neutron-capture process; Paper~III).

This paper is structured as follows: Section 2 describes the sample of
$r$-process-enhanced stars explored in our programme. Sections 3 and 4
describe in detail the observations, and the reduction and analysis
procedures employed in the papers of this series. Section 5 summarizes
our results, while Section 6 discusses these results and their
implications for the progenitors of the $r$-I and $r$-II stars and the
associated enrichment processes. Section 7 presents our final
conclusions.


\section{Sample stars}

All but two of our sample stars were selected from the HK survey of
Beers and colleagues \citep{beers1985,beers1992} and the Hamburg/ESO
survey of Christlieb and collaborators \citep{christlieb2008}. The
sample includes a number of the canonical examples of the
$r$-process-element enhancement phenomenon; \object{CS~22892$-$052}, the
first EMP star with detected Th \citep{mcwilliam1995,sneden2000},
\object{CS~31082$-$001}, the first EMP star with detected U \citep{hill2002,
cayrel2004}, and \object{HE~1523$-$0901}, the most extremely $r$-process
enhanced giant known \citep{frebel2007}. All except HE~1523$-$0901 were
analysed in the Hamburg/ESO R-process Enhanced Star (HERES) survey
\citep{christlieb2004,barklem2005}. Accordingly, most of our programme
stars are in the Southern Hemisphere (but north of declination $\delta
\sim -25\degr$) and have $V\lesssim16$, which is the practical limit for
1-hour integrations with the NOT. 

The sample stars are listed in Table \ref{tbl-1}, including their $V$ magnitudes 
and $B-V$ colours, and reported $\mathrm{[Fe/H]}$ and $\mathrm{[Eu/Fe]}$
abundances. The last column of this table indicates whether a given star
is considered to be a member of the moderately $r$-process-enhanced
class ($r$-I; $+0.3  \le \mathrm{[r/Fe]} \le +1.0$) or the highly
$r$-process-enhanced class ($r$-II; $\mathrm{[r/Fe]} > +1.0$), according
to the definitions of \citet{beerschristlieb2005}. 

\section{Observations}

Spectra for our radial-velocity monitoring programme were obtained with
the NOT, in service mode, using the FIES
spectrograph\footnote{http://www.not.iac.es/instruments/fies/}, which
has been used successfully for exo-planet research
\citep{buchhave2012}. The spectra cover the wavelength range 3640\,{\AA} $-$
7360\,{\AA} in 78 orders, at a resolving power of $R~\sim$46,000. The S/N of
the spectra is $\sim$10 on average, but ranges from $\sim$2 to $\sim$20. A S/N
of $\sim$10 is obtained in $\sim$20 min for a star of $V = 14.5$, so a typical
clear night yielded $\sim$10-15 spectra of the stars in Table
\ref{tbl-1}. Integrations of 900~s or longer were split into three
exposures, in order to enable effective cosmic ray rejection. 

The observing strategy was based on the assumed analogy with the Ba~II and CH
binaries found by \cite{mcclure1984} and \cite{mcclure1980} to have periods of
the order of $\sim$ 300-3000 days and amplitudes of $\sim$ 3-10  km~s$^{-1}$. Accordingly,
spectra have been obtained at roughly monthly intervals since June 2007, and
reduced immediately, so that follow-up of any variable objects could be planned
efficiently. As we demonstrate, this strategy has worked well. 

\section{Reduction and analysis}

The observations were reduced with pipeline software originally
developed by Lars Buchhave to deliver high-precision radial velocities
of exo-planet host stars from echelle spectrographs, in particular the
FIES instrument \citep{buchhave2010}. 

This reduction procedure includes all the normal steps, such as bias
subtraction, division by a flat-field exposure, cosmic ray removal, and 2-D
order extraction. Sky background is only significant for the
faintest stars if observed close to the full Moon, which was avoided in
the nightly planning. Moreover, even substantial amounts of scattered
moonlight are harmless if the two cross-correlation peaks are well
separated, which is virtually always the case for our high-velocity
programme stars.

For the wavelength calibration, a separate wavelength
solution is created for each target spectrum, using Th-Ar calibration
spectra taken just before each science frame. This procedure has been
found to yield adequate velocity stability, as we demonstrate below. 

\subsection{Multi-order cross-correlation}

With the reduced spectra in hand, multi-order cross-correlation against
an optimized template spectrum is then performed, using software also
developed by L. Buchhave \citep{buchhave2010}. The radial velocity from
each individual order is determined by a Gaussian fit to the peak of the
cross-correlation function (CCF); their mean value, weighted by the
total photon count in each order, is taken as the final radial velocity from
the observation. 

Performing the cross-correlation order-by-order enables us to hand-pick the
spectral regions to be used in the correlation, including regions with strong
absorption lines and excluding regions with only a few and/or weak lines, which is
a significant advantage when dealing with spectra of stars of such peculiar
chemical compositions. Filtering is also applied to the spectrum before the
cross-correlation to remove unwanted frequencies. The filters are carefully
optimized for each star to remove noise while retaining even the narrowest
stellar absorption lines.

\subsection{Optimization of the template spectra}

The choice of the template spectrum for the cross-correlation is crucial for the
accuracy of the resulting radial velocities, especially in these spectra where
the usual iron-peak elements only show weak lines, but strong lines exist
from the normally rare neutron-capture elements.

Four different recipes have been used to construct the template spectra,
depending on the quality of the target spectra: ``Strongest'', ``Co-add'',
CS~31082$-$001, and ``Delta''. The Strongest template is the spectrum of
a given star with the maximum signal level for that star. The advantage
of using a spectrum of the same star as a template is the perfect match
to the observed spectrum; the disadvantage is that a template with
relatively low S/N will introduce noise into the correlation. The Co-add
template is constructed by shifting a selection of the best spectra of the
star to a common radial velocity and co-adding them. This results in a
template with high S/N (which is also a perfect match for the target spectra),
and will generally allow more orders to be included in the correlation,
compared to a correlation with the strongest single spectrum as
template. However, when creating the Co-add template, an initial correlation
with the strongest spectrum as template is used to determine the shift of the
other spectra. Any small residual shift will then broaden the spectral lines
in the Co-add template spectrum and yield less precise results in the final
correlation. 

For the fainter stars, the Strongest template may introduce too much
noise into the correlation, and a good Co-add template cannot then
be constructed. Instead, a Co-add template from a bright star with a
very similar spectrum may then be used as template; for these stars we
have used CS~31082$-$001 in this manner. The final template option, used
here for the faintest stars, is the Delta template, a synthetic
spectrum consisting of $\delta$ functions at the (solar) wavelengths of
selected lines. Sample spectra of the faint stars have been inspected
for strong lines to be included in the Delta template spectrum.
Correlation with this template yields velocities on an absolute scale,
and has thus also been used to determine the absolute velocity of the
other templates and to convert all velocities to an absolute scale. 

\begin{table*}
\caption{Results for the observed radial-velocity standard stars}
\label{tbl-2}
\centering
\begin{tabular}{lrrrrrcrr}
\hline\hline
Stellar ID & RA (J2000) & Dec (J2000) & $B$ & $V$ & RV mean & $\sigma$ & Nobs & $\Delta$T\\
           &            &             &     &     & (km~s$^{-1}$)&(km~s$^{-1}$)  & &(days) \\
\hline
\object{HD~3765}   & 00:40:49 & $+$40:11:14 & 8.30 & 7.36 &$-$63.35 & 0.033 & 60 & 2672\\
\object{HD~38230}  & 05:46:02 & $+$37:17:05 & 8.19 & 7.36 &$-$29.14 & 0.035 & 46 & 2771\\
\object{HD~79210}  & 09:14:23 & $+$52:41:12 & 9.07 & 7.63 &$+$10.45 & 0.048 & 29 & 2746\\
\object{HD~115404} & 13:16:51 & $+$17:01:02 & 7.46 & 6.52 & $+$7.67 & 0.036 & 34 & 2891\\
\object{HD~151541} & 16:42:39 & $+$68:11:18 & 8.32 & 7.56 & $+$9.44 & 0.028 & 30 & 2857\\
\object{HD~182488} & 19:23:34 & $+$33:13:19 & 7.15 & 6.36 &$-$21.65 & 0.035 & 61 & 2072\\
\object{HD~197076} & 20:40:45 & $+$19:56:08 & 7.06 & 6.44 &$-$35.37 & 0.052 & 26 & 2857\\
\hline
\end{tabular}
\end{table*}

\begin{table*}
\caption{Mean heliocentric radial velocities, standard deviations, and
  time-span covered for the programme stars} 
\label{tbl-3}
\centering
\begin{tabular}{lrlrrcc}
\hline\hline
Stellar ID & Nobs &  Template & RV mean    & $\sigma$ & $\Delta$T &Binary\\
           &      &           &(km~s$^{-1}$)&(km~s$^{-1}$)& (days) & \\
\hline
\object{HD~20}         &14 & Strongest      & $-$57.914 & 0.041 & 2603 & No\\
\object{CS~29497$-$004}&12 & Co-add         &$+$105.008 & 0.366 & 2583 & No\\
\object{CS~31082$-$001}&24 & Co-add         &$+$139.068 & 0.105 & 2642 & No\\
\object{HE~0432$-$0923}&18 & Delta          & $-$64.800 & 0.988 & 2737 & No\\
\object{HE~0442$-$1234}&28 & Co-add         &$+$237.805 & 8.294 & 2618 & Yes\\
\object{HE~0524$-$2055}&13 & CS~31082$-$001 &$+$255.425 & 0.195 & 2338 & No\\
\object{HE~1044$-$2509}&14 & Delta          &$+$365.789 &17.110 & 1887 & Yes \\
\object{HE~1105$+$0027}& 9 & Delta          & $+$76.197 & 0.496 & 1573 & No\\
\object{HE~1127$-$1143}& 7 & Delta          &$+$229.157 & 0.454 & 1998 & No\\
\object{HE~1219$-$0312}& 5 & Delta          &$+$162.416 & 1.094 & 2171 & No\\
\object{HE~1430$+$0053}&20 & Co-add         &$-$107.749 & 0.426 & 2493 & No\\
\object{HE~1523$-$0901}&34 & Co-add         &$-$163.271 & 0.284 & 2594 & Yes\\
\object{CS~22892$-$052}&19 & Co-add         & $+$13.549 & 0.164 & 2174 & No\\
\object{HE~2224$+$0143}&24 & Co-add         &$-$113.085 & 0.190 & 2420 & No\\
\object{HE~2244$-$1503}&14 & Delta          &$+$147.928 & 0.246 & 2207 & No\\
\object{HD~221170}     &30 & Strongest      &$-$121.201 & 0.105 & 2174 & No\\
\object{CS~30315$-$029}&14 & Co-add         &$-$169.346 & 0.352 & 2672 & No\\
\hline
\end{tabular}
\end{table*}

\subsection{Standard stars} 

A few well-established radial-velocity standard stars from Table 2 of
\cite{udry1999} were observed on every usable observing night in order to
monitor any zero-point variations in the velocities. These are listed in
Table~\ref{tbl-2}, along with our mean (heliocentric) velocities and dispersions for these
stars. The mean difference of our velocities from the standard values is 73
m~s$^{-1}$, with a standard deviation of 69 m~s$^{-1}$. This demonstrates
that our results for the target stars are on a system consistent with that 
of \cite{udry1999}, and that the accuracy of our results is not limited by 
the stability of the spectrograph.

\subsection{Error estimates} 

The internal error on the mean velocity from each spectrum is calculated 
as the standard deviation of the velocities from the individual orders used in 
the correlation. It is listed along with each observed velocity in Appendix A, 
and is used to plot the error bars of the data in the orbital plots shown in
Fig. \ref{fig:orbit}. For each star, the average internal error on the
velocities is computed as the mean of these internal standard deviations, and
is used to assess the quality of the correlation and the order selection. 

The {\it external} standard deviation, $\sigma$, of the radial-velocity
observations for each star, given in Table \ref{tbl-2} for the radial-velocity
standards and in Table \ref{tbl-3} for the programme stars, is computed as 

{\centering
\begin{eqnarray}
\sigma = \sqrt{\frac{1}{N-1}\sum^{N-1}_{j-0}(v_j-\bar{v})^2}.
\end{eqnarray}

}
\noindent Imperfect guiding and centring of the star on the fibre end along
with imperfect cancellation of changes in ambient temperature and atmospheric pressure,
etc., contributes to small variations in the derived radial velocities,
which are reflected in the standard deviations of the constant programme
and radial-velocity standard stars.

For the three detected binaries in our sample, the above value for $\sigma$ is
inflated due to the orbital motion, and the relevant uncertainty estimate is the
$\sigma$ from the orbital solutions given in Table \ref{tbl-4}. 

\section{Results}

\begin{table*}
\caption{Orbital parameters of the binary systems in the sample}
\label{tbl-4}
\centering
\begin{tabular}{lccc}
\hline\hline
Parameter & \object{HE~0442$-$1234} & \object{HE~1044$-$2509} & \object{HE~1523$-$0901} \\
\hline
Period (days)        &2515.2$\pm$5.5   & 36.561$\pm$0.009 & 303.05$\pm$0.25\\
$T_0$ (BJD)          &2457918.3$\pm$1.0& 2455737.1$\pm$0.1& 2455068.0$\pm$4.1\\
$K$ (km~s$^{-1}$)     &12.541$\pm$0.016 & 27.024$\pm$0.96    & 0.350$\pm$0.003\\
$\gamma$ (km~s$^{-1}$)& 236.35$\pm$0.01 & 360.22$\pm$0.57  &-163.23$\pm$0.003\\
$e$                  & 0.767$\pm$0.001 & 0.000\tablefootmark{a}             & 0.163$\pm$0.010\\
$\omega$~$\degr$      & 316.6$\pm$0.1   & 90.00               & 81.7$\pm$4.7\\
$a\sin$ $i$ ($R_{\sun}$)& 400.2$\pm$1.2  & 19.6$\pm$0.7      & 2.07$\pm$0.02 \\
$f(m)$ ($M_{\sun}$)      & 0.136$\pm$0.001& 0.075$\pm$0.009   & 1.3E-5$\pm$4E-8 \\
$\sigma$ (km~s$^{-1}$)& 0.28           & 1.88               & 0.11\\
$R_{Roche}$ ($R_{\sun}$, $M_1$ = 0.8 $M_{\sun}$, $M_2$ = 0.6 $M_{\sun}$) & 59 & 13 & 53 \\
$R_{Roche}$ ($R_{\sun}$, $M_1$ = 0.8 $M_{\sun}$, $M_2$ = 1.4 $M_{\sun}$) & 770 & 46 & 220 \\
\hline
\end{tabular}
\tablefoot{
\tablefoottext{a}{Eccentricity for HE1044-2509 fixed to zero.}}
\end{table*}

The results of our radial-velocity monitoring of the sample stars are
summarized in Table \ref{tbl-3}, which lists the star name, the number
of observations (Nobs), the template used for each star, the mean
(heliocentric) radial velocity and standard deviation over the observed
time span ($\Delta$T), and the binary status for each star. The
individual observed heliocentric radial velocities are listed in
Appendix A, together with the Julian dates of the observations and their
internal errors. 

As can be seen from a glance at Table \ref{tbl-3}, fourteen of our stars
exhibit no variation in their radial velocities at the level of a few
hundred metres per second over the eight years of monitoring. For the
brighter targets, the standard deviations of the observed velocities are
$\sim$100 m~s$^{-1}$ (dominated by centring and guiding errors), rising to 
$\sim 1$ km~s$^{-1}$ for the fainter targets, due to the lower S/N of their 
spectra. Moreover, least-squares fits of the velocities vs. time reveal no 
net trends over the observing period.

\subsection{Comparison with literature data}

Table \ref{tbl-B1} in appendix B lists mean radial velocities
(based on high-resolution spectroscopy) from
the literature for the single stars in our sample, along with the
complete time span covered by the combined data (including our own
measurements). Most of the stars have only been observed once earlier
by \citet{barklem2005}, for which they estimate an error of a few km~s$^{-1}$,
and we find excellent agreement between their results and ours. For HD~20
and  HD~221170, radial velocities were also reported by \citet{carney2003},
who list  13 observations for HD~20 and 18 for HD~221170, spanning 4641 and
5145 days, with (external) standard deviations of 0.41 and 0.61
km~s$^{-1}$, respectively. Their mean radial velocities of $-57.18$
km~s$^{-1}$ and $-121.77$ km~s$^{-1}$ are consistent with our results,
given the slight offset of the CfA velocities from the system of
\citet{udry1999}. 

For the faint star HE~1219$-$0312, our five measurements with the 2.5-m
NOT span 2171 days from 2007, while the five epochs of the earlier
\citet{hayek2009} VLT/UVES observations span $\sim400$ days. The
standard deviations ($\sim$1 km~s$^{-1}$) and their mean radial velocity
(163.1 km~s$^{-1}$) are similar to ours, and consistent with our own
conclusion that this star is single.  

From inspection of Table \ref{tbl-B1} it is clear that, while the
coverage of the listed spans for most of the single stars is rather
sparse, the spans themselves are from two to five times longer than
those obtained during the course of our own radial-velocity monitoring
programme. The lack of observed variations beyond what can be accounted
for by the expected errors across multiple spectrograph/telescope
combinations strengthens our claim that the stars we classify as single
are indeed so.

\subsection{Binary orbits}

Three of our stars are spectroscopic binaries, as first reported by
\citet{hansen2011}. HE~0442$-$1234 was found already by P. Bonifacio et
al. (private comm.) to be a long-period binary, while HE~1044$-$2509 and
HE~1523$-$0901 are new discoveries from our programme. All three orbits
have now been fully completed, with particular attention being paid to
assessing the reality of the very low-amplitude velocity variations of
HE~1523$-$0901. After rejecting early observations made under poor
conditions (strong moonlight, poor seeing), and with now almost nine
orbital revolutions completed, we have satisfied ourselves that this
nearly face-on orbit is real and the orbital parameters are reliable
(see below). 

\begin{figure}
\resizebox{\hsize}{!}{\includegraphics{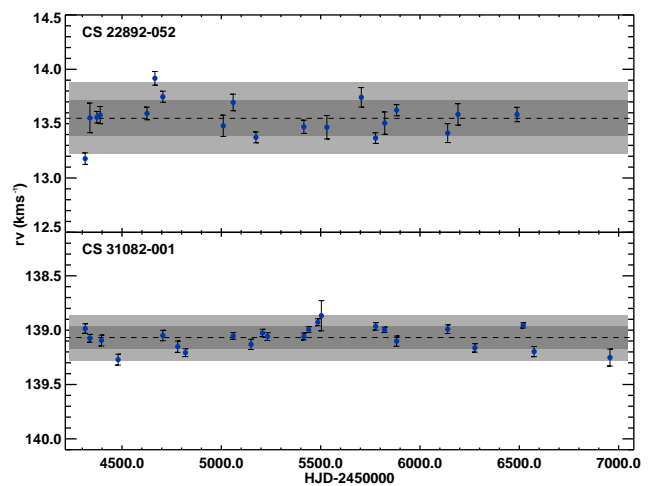}}
\caption{Heliocentric radial velocities derived for CS~22892$-$052 (top) and CS~31082$-$001 (bottom), as a function of time.Dashed lines: Mean radial velocity; grey shaded areas: 1$\sigma$ and 2$\sigma$ regions
around the mean.} 
\label{fig:single}
\end{figure}

In contrast to this result, our own (far more accurate) velocity data 
clearly disprove the putative orbit suggested by 
\cite{prestonsneden2001} for CS~22892-052, with $P \sim~128$ days and $K
\sim~1.0$ km~s$^{-1}$. Both this star and CS~31082$-$001 are clearly
single stars, as can be seen from Table \ref{tbl-3} and Figure 
\ref{fig:single}, where the derived radial velocities for the two stars
are plotted as a function of time. Our results for stars observed with similar 
time spans and accuracies in parallel parts of our overall programme show 
that binaries with orbital periods of 20-30 years are detected with certainty 
even after the first couple of  years with $\sim$monthly observations.

The observed $K \sim350$ m~s$^{-1}$ of HE~1523$-$0901 is the smallest
measured with certainty in our programme, highlighting the null result
for the other, single stars. The corresponding tiny mass function
implies either a companion of mass in the brown-dwarf range (for $i \sim
90\degr$), or a very low orbital inclination, or a combination of both.
Assuming $i \la2.5\degr$ leads to a secondary mass in the late M-dwarf
range, 0.25 $M_\sun$ -- still plausible within the statistics for a
single case in a sample of several tens of potential (southern) $r$-I
and $r$-II targets. 

Long-period envelope pulsations are a potential alternative origin of 
radial-velocity variations for this and other cool giant stars, and we have 
also found some strongly carbon-enhanced VMP and EMP stars showing similar low-level 
radial-velocity fluctuations. The frequency of real spectroscopic 
binaries with very low inclinations is far too low to ascribe all such 
low-amplitude velocity variations to binary orbital motion, and most marginal
variations are, in the end, found not to be strictly periodic. Thus, it may be
that low-level pulsations, rather than velocity accuracy, may set the ultimate
limit to the length of the binary periods that can be reliably detected by the
radial-velocity technique, as is the case for exo-planet orbits.

Our current knowledge of the pulsational characteristics of late-type evolved 
stars is derived from the systematic microlensing surveys of MACHOs towards 
the Magellanic Clouds \citep{wood2000,riebel2010}. The late-type pulsators with 
periods of 200-1000 days are typically C-type AGB stars or Mira variables, but
stars below the tip of the red-giant branch have shorter periods, and the cause of their 
light variation is not known. Precise light and colour curves for field
VMP/EMP stars with equally well-determined distances, or at least log $q$, periods or
pseudo-periods of the order of a year, and phasing consistent with pulsations would
be needed to settle the issue definitively. However, obtaining them with
sufficient accuracy is not an easy task, and none has been reported for
HE~1523$-$0901. 

\begin{figure}
\resizebox{\hsize}{!}{\includegraphics{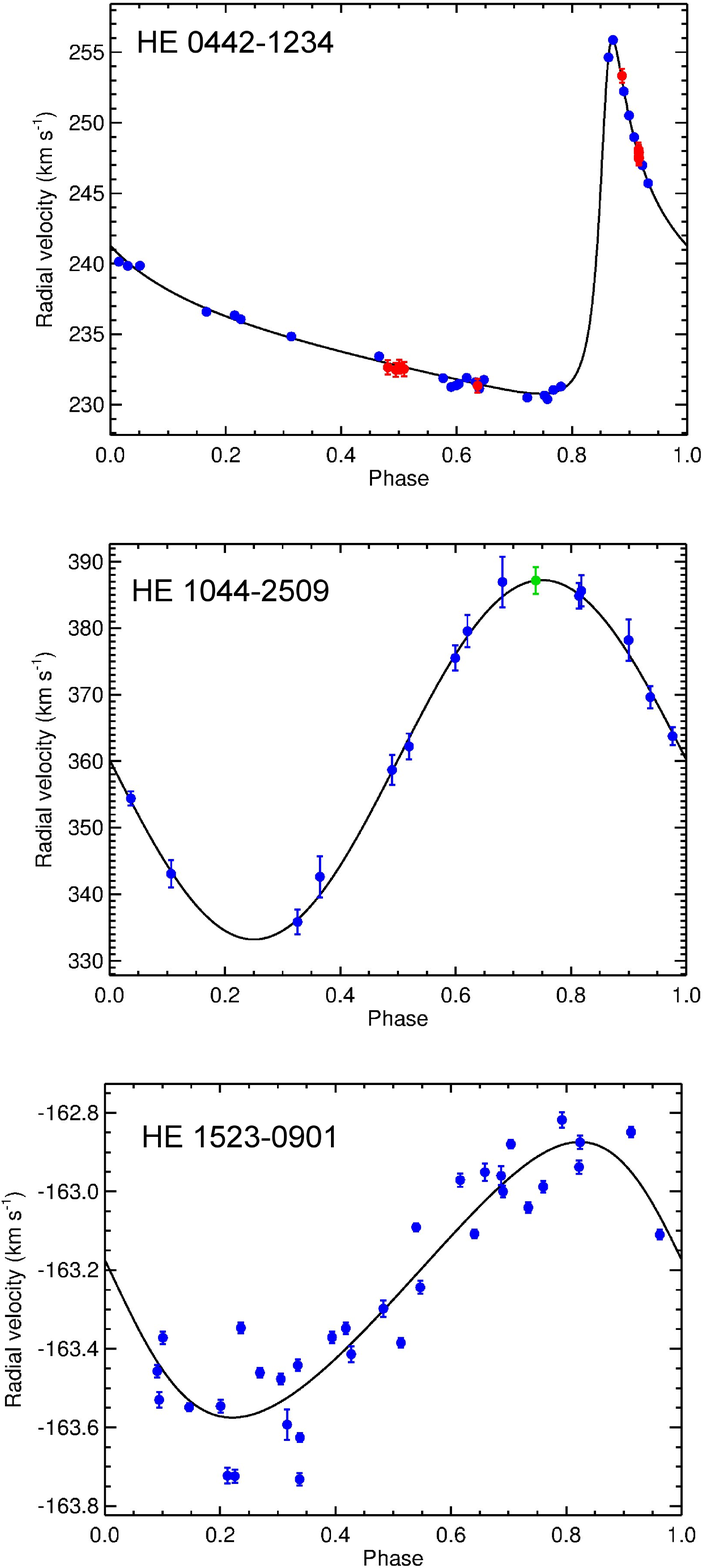}}
\caption{Orbital solutions for the three binary systems found in our programme.
Top: HE~0442$-$1234 (blue this work, red P. Bonifacio private comm.),
  middle: HE~1044$-$2509 (blue this work, green \citet{barklem2005}) and bottom:
  HE~1523$-$0901.} 
\label{fig:orbit}
\end{figure}

The final orbital elements for the three binary systems among our
programme stars are listed in Table \ref{tbl-4}, and velocity curves with
all available observations are shown in Figure \ref{fig:orbit}.

\section{Discussion}

\subsection{Binary frequency of $r$-I and $r$-II stars}
\label{discuss:binary}

The salient result of our study is that only 3 of our 17 programme stars are
binaries, while 14 are confirmed to be single stars, a binary frequency of 
$\sim18\pm$6\%. 

Our sample is relatively small, since HERES stars south of the NOT limit 
could not be observed by us, but only two other $r$-process-enhanced stars, 
the $r$-II star \object{HE~2327$-$5642} and the $r$-I star
\object{CS~22183$-$031}, have been reported in the literature to show variable
radial velocities. \object{HE~2327$-$5642} was discovered by
\citet{mashonkina2010}; their data cover a range of $\sim$4.3 years,
during which the radial velocity of the star varied by $\sim$20
km~s$^{-1}$. The other star, \object{CS~22183$-$031}, is included in the
sample of \citet{roederer2014b}, who reported on the identification of nine
new $r$-process-enhanced stars. The radial-velocity data for neither of these
stars is sufficient to derive an orbital solution for the systems. The bright
$r$-I star \object{HD~115444} \citep{westin2000} was not observed in our
programme, but literature data confirm (with variations of no more than 1
km~s$^{-1}$ over a sparsely-sampled range of 24 years; see Table \ref{tbl-B2})
that it too is vey likely a single star, as is its $r$-process-{\it poor}
counterpart \object{HD~122563} (with variations of no more than 1.5
km~s$^{-1}$ over a sparsely-sampled range of 59 years; see Table
\ref{tbl-B2}). 

In summary, the binary frequency of 18$\pm$6\% found here for the
$r$-I and $r$-II stars is completely consistent with the 16$\pm$4\% of
binaries with periods up to 6,000 days found by \cite{carney2003} in
their survey of 91 metal-poor field giants, and the $\sim$22\%
binaries with periods up to $\sim$15,000 days found by
\cite{mermio2008} in their sample of $\sim$1,300 Population I giants in
Galactic open clusters. A binary frequency of 100\% for this class of
VMP and EMP stars is clearly ruled out, a conclusion that would
only be reinforced if HE~1523$-$0901 were eventually proved to be a
single pulsating star rather than a binary with a nearly face-on orbit. 

Overall, we can thus conclude that the observed dramatic excess of
$r$-process elements in our sample of stars is not just a surface effect
produced locally by a binary companion, but rather was produced by a
remote source and imprinted on the parent clouds across interstellar
distances. It also seems hard to imagine how the local binary scenario
could produce stars like \object{HD~122653}, with a {\it deficit} of
$r$-process elements relative to the standard abundance pattern. 

\begin{figure}
\resizebox{\hsize}{!}{\includegraphics{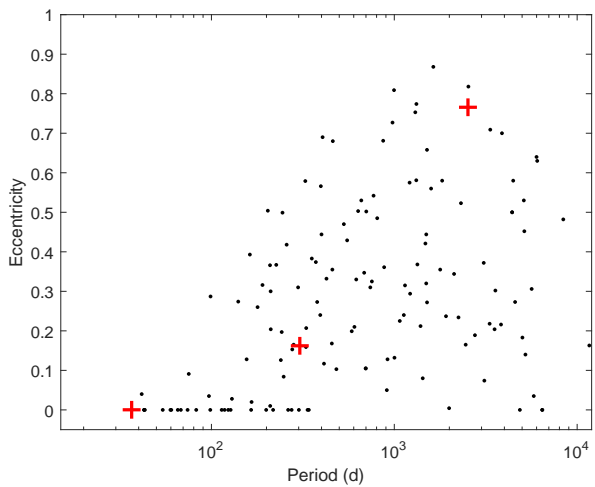}}
\caption{Period -- eccentricity diagram for giant binaries. Dots: 141 members of
  Galactic open clusters \citep{mermio2007,mathieu1990}; red plus signs: The three
  $r$-I and $r$-II binaries discussed in this paper.} 
\label{Pe_plot_r}
\end{figure}

\subsection{Frequency and properties of $r$-I and $r$-II binaries}
\label{discuss:orbits}

The distribution of periods and eccentricities of our three binaries
is also fully as expected for binary systems with normal giant
primary stars. This is illustrated most simply in Figure
\ref{Pe_plot_r}, which shows a period -- eccentricity diagram
constructed from the orbital data for 141 giant binary members of
(Population I) Galactic open clusters by \citet{mermio2007} and
\citet{mathieu1990}, plotted as dots, while our three binaries are shown as
red plus symbols. As seen, their orbital eccentricities are completely
normal for giant binaries, which are typically tidally circularized for
periods up to $\sim150$ days, depending on age and stellar mass, and
show no sign of tidal or other processes that could be connected to
their outstanding chemical peculiarity. In contrast, the transition
between circular and eccentric orbits seems to occur at periods of
$\sim4-800$ days for binaries with former AGB companions. 

Further information on the presently unseen companions can be derived by
considering the volume available to them during their evolution; i.e.
their Roche-lobe radii, which can be computed from the observed orbital
separation (assuming no exchange or loss of mass and angular momentum)
and a range of assumed stellar masses. We first adopt a common
mass of 0.8~$M_\sun$ for the observed EMP giants and assume that the
companion is a (sub)dwarf (i.e. unevolved) star of mass 0.6~$M_\sun$, at
least three magnitudes fainter than the star we do see. The present radius of
the latter can be estimated from the adopted mass and a typical $\log~g
\sim1.5$ dex, i.e. R $\sim$ 30 $R_\sun$. 

With the observed orbital elements, notably $a_1$sin$i$ and the mass
function, we then adjust $i$ until the computed $M_2$ reaches 0.6
$M_\sun$, and calculate the corresponding Roche-lobe radii, which are
given in Table \ref{tbl-4}. For HE~0442$-$1234, the minimum secondary
mass is 0.67 $M_\sun$ already for $i$ = 90$\degr$; for HE~1044$-$2509
and HE~1523$-$0901, we find $i$ = 62$\degr$ and 1.3$\degr$,
respectively, for a secondary mass of 0.6~$M_\sun$. 

However, we might alternatively assume that the companions were
initially {\it more} massive than 0.8~$M_\sun$, i.e. in the range of
1-8~$M_\sun$, where they would have gone through the AGB phase and
likely have evolved into now-invisible white dwarfs (WDs) with typical
masses of 0.6~$M_\sun$, and thus identical R$_{Roche}$ to the ones found
above. It is possible that the WD might also have the maximum WD mass
of 1.4~$M_\sun$, in which case the entire system would be much larger
and the Roche-lobe radius of the companion a larger share of that again.
Nominal Roche-lobe radii for this case are also given in Table
\ref{tbl-4}, and illustrate the dramatic change in the space available
to the star during its evolution, depending on its assumed mass. 

For HE~1044$-$2509, even this R$_{Roche}$ is still too small to
accommodate a typical AGB star of $\sim$200 $R_\sun$. In any case, the
observed absence of any $s$-process signatures indicates that mass
transfer from a putative AGB companion did not happen in any of these
systems, while a supernova explosion of a putative, even more massive
companion would likely have disrupted the binary system \citep{tauris1998}.

In summary, binary systems seem to occur as a normal part of the
formation of $r$-process-enhanced metal-poor stars, and are unrelated to
the process by which they acquired their outstanding chemical anomalies.

\subsection{Origin of $r$-I and $r$-II stars}

The separation of $r$-process-enhanced stars into the $r$-I and $r$-II
classes was originally a matter of convenience, since the heavy-element
abundance patterns of the most enhanced stars could be more easily
observed due to their relatively stronger lines at low metallicity. As
the sample of such stars has grown, it has become clear that these
classes also exhibit rather different behaviour with metallicity; the
$r$-II stars are found in a relatively narrow range of metallicity near
$\mathrm{[Fe/H]} \sim -3.0$, while the $r$-I stars cover a larger range
of metallicity, $-3.5 < \mathrm{[Fe/H]} < -1.5$ \citep{beers2013}. It
remains unclear whether this constraint implies that different classes
of astrophysical progenitors might be responsible for the $r$-I and
$r$-II stars, or whether the initial $r$-process content in the natal
clouds of the $r$-I stars has simply been diluted by the elemental mix
of standard chemical evolution. 

All of our programme stars are giants. However, it is important to note
that \citet{aoki2010} have shown that the star
\object{SDSS~J2357$-$0052} is a cool (T$_{\rm{eff}} \sim $5000~K)
main-sequence dwarf with $\mathrm{[Fe/H]} = -3.4$, and $\mathrm{[Eu/Fe]}
= +2.0$, making it simultaneously the lowest metallicity and most
Eu-enriched $r$-II star yet found. This star is of particular interest,
as all previous $r$-process-enhanced stars identified to date have been
in more evolved stages of evolution. Since dwarfs with the temperature
of SDSS~J2357$-$0052 do not have convective atmospheres, it can be
reasonably concluded that the $r$-process-enhancement phenomenon is not
due to some chemical peculiarity arising from the presence of a
convective envelope in such stars. Unfortunately, although this star is
sufficiently bright ($V \sim $15.6) for the programme described above,
it was discovered too late to be included in our target list, but the
results of radial-velocity monitoring over about a year by Aoki et al.
did not reveal any evidence of significant variation. 

The nine newly recognized $r$-process-enhanced stars (based on
high-resolution spectroscopic follow-up of HK survey stars) by
\citet{roederer2014b} include subgiants and the field equivalents of red
horizontal-branch stars, reaffirming that $r$-process enhancement is
not an evolutionary effect in the stars. \citet{roederer2014b} also
compare the abundance pattern of the light elements in their
$r$-process-enhanced stars with that of {\it non-} $r$-process-enhanced stars
with similar stellar parameters. No evidence was found to indicate that
the $r$-process-enhanced stars have different abundance patterns for the
light elements than for the comparison sample, leading the authors to
conclude that the event(s) producing the high levels of $r$-process
material seen in these stars do not produce a distinct light-element
abundance pattern. Neither does the large $r$-process enhancement seem
to be coupled to the carbon and nitrogen abundances in the stars,
although a few of the $r$-II stars are found to also be enhanced in
carbon, the most well-known example being CS~22892$-$052 \citep{sneden2000}.  

\citet{mashonkina2010} explored the Sr, Ba and Eu abundances for a number of $r$-I and
$r$-II stars. They found very similar $\mathrm{[Ba/Eu]}$ abundance
ratios for the two groups ($\mathrm{[Ba/Eu]} \sim +0.60$), but the mean
$\mathrm{[Sr/Eu]}$ ratio differed by 0.36 dex between the two groups,
with the $r$-II stars having the lowest ratio, $\mathrm{[Sr/Eu]} =
-0.93$. The authors argued that elements from the first and second
$r$-process peak are of common origin in the $r$-II stars, whereas for
the $r$-I stars the picture is less clear.

Competing scenarios for the origin of the $r$-I and $r$-II stars invoke
non-spherical, jet-producing supernova explosions or neutrino winds from
merging neutron star binaries. Detailed observations of the abundance
patterns predicted from the two competing scenarios may be the
best guide to identifying the production site(s), but the frequency of
these stars ($\sim3\%$ among VMP and EMP stars; \citealt{roederer2014b})
provides another clue; any jets in such scenarios must be highly
collimated in order to selectively enrich only a small fraction of
molecular clouds in the early ISM.  

One circumstance is worthy of note, although its interpretation is
currently unclear: Of the two $r$-II stars with secure detections of
uranium, \object{CS~31082$-$001} exhibits the so-called ``actinide
boost'' \citep{hill2002}, and has now been shown to be a single star,
while the detailed $r$-process abundance pattern of the newly identified
binary \object{HE~1523$-$0901} does not. While this may just be a result
of small-number statistics on these very rare objects, it should be kept
in mind as the sample grows; the details of their chemical-abundance
patterns would appear to offer our most reliable clue to the origin of
this difference.

\section{Conclusions}

In order to detect a possible link between the nature of the binary
population among stars that exhibit moderate to strong enhancements in
their $r$-process-element abundances and the origin of their peculiar
abundance patterns, we have monitored the radial velocities of 17 such stars
with high precision ($\sim100$ m~s$^{-1}$) over a period of eight years. 

Of the 17 programme stars, 14 exhibit no radial-velocity
variations during this period, and are thus presumably single, while
three are binaries with normal orbital periods and eccentricities,
yielding a normal binary frequency of $\sim$18\% among these stars.
Hence, there is no evidence that the $r$-process enhancement seen in the
$r$-I and $r$-II stars is causally linked to the binary nature of the
stars. Furthermore, the detection of $r$-process-element rich stars in
dwarf, subgiant, giant, and horizontal-branch evolutionary phases shows
that the enhancement phenomenon is also not linked to the evolution of
the stars. 

We conclude that the moderate-to-high $r$-process-element abundances
derived for these stars must be an imprint of the clouds from which the
stars were formed. Detailed abundance analysis of larger samples of
$r$-I and $r$-II stars is required to better constrain the nature of the
objects and the chain of events that polluted the natal clouds of these
stars. Moreover, the mechanism(s) for such sporadic, inhomogeneous
enrichment of the ISM in early galaxies, and their implication for
understanding the observed chemical composition of high-redshift DLA
systems, should be considered in the next generation of models for
the formation and early chemical evolution of galaxies.

\begin{acknowledgements}

This paper is based on observations made with the Nordic Optical
Telescope, operated by the Nordic Optical Telescope Scientific
Association at the Observatorio del Roque de los Muchachos, La Palma,
Spain, of the Instituto de Astrofisica de Canarias. 

We thank several NOT staff members and students for obtaining most of the
observations for us in service mode. 

The work of T.H. was supported by
Sonderforschungsbereich SFB 881 ``The Milky Way System'' (subproject A4)
of the German Research Foundation (DFG). T.C.B. and J.Y. acknowledge partial
support for this work from grants PHY 08-22648; Physics Frontier
Center/{}Joint Institute or Nuclear Astrophysics (JINA), and PHY
14-30152; Physics Frontier Center/{}JINA Center for the Evolution of the
Elements (JINA-CEE), awarded by the US National Science Foundation. J.A.
and B.N. gratefully acknowledge financial support from the Danish
Natural Science Research Council and the Carlsberg Foundation. 
B.N. gratefully acknowledges partial support from the National Science 
Foundation under Grant No. NSF PHY11-25915.

Furthermore, we thank the referee for helpful comments.

\end{acknowledgements}


\clearpage

\begin{appendix}
\section{Individual heliocentric radial velocities measured for the
programme stars}

\begin{table}[ht]                   
\caption{HD~20}         
\centering                       
\begin{tabular}{lcr}
\hline\hline
HJD & RV & RV$_{err}$  \\
    & (km~s$^{-1}$) & (km~s$^{-1}$)\\
\hline
2454314.661653&  $-$57.892&  0.033\\
2454338.615723&  $-$57.891&  0.022\\
2454373.612652&  $-$57.900&  0.024\\
2454819.293270&  $-$57.983&  0.030\\
2455126.580295&  $-$57.946&  0.041\\
2455175.394755&  $-$57.978&  0.023\\
2455503.480513&  $-$57.855&  0.070\\
2455776.652777&  $-$57.940&  0.022\\
2455796.657395&  $-$57.927&  0.025\\
2455859.514789&  $-$57.862&  0.034\\
2456139.703103&  $-$57.863&  0.027\\
2456241.446457&  $-$57.899&  0.064\\
2456529.679820&  $-$57.934&  0.023\\
2456917.579578&  $-$57.924&  0.027\\
\hline
\end{tabular}           
\end{table}

\begin{table}[ht]                   
\caption{CS~29497$-$004}         
\centering                       
\begin{tabular}{lcr}
\hline\hline
HJD & RV & RV$_{err}$  \\
    & (km~s$^{-1}$) & (km~s$^{-1}$)\\
\hline
2454373.642892&  104.329&  0.116\\
2454705.631179&  104.873&  0.065\\
2454780.500993&  105.409&  0.292\\
2454819.320354&  105.561&  0.084\\
2455175.419215&  105.010&  0.063\\
2455415.686599&  104.941&  0.061\\
2455439.606743&  105.124&  0.060\\
2455796.671082&  105.434&  0.077\\
2455858.545060&  105.060&  0.060\\
2456191.589163&  105.050&  0.045\\
2456530.687786&  104.829&  0.054\\
2456956.535485&  104.471&  0.094\\
\hline                  
\end{tabular}           
\end{table}

\begin{table}[ht]                   
\caption{CS~31082$-$001}         
\centering                       
\begin{tabular}{lcr}
\hline\hline
HJD & RV & RV$_{err}$  \\
    & (km~s$^{-1}$) & (km~s$^{-1}$)\\
\hline
2454314.681742&  138.985&  0.044\\
2454338.656269&  139.075&  0.036\\
2454396.548819&  139.095&  0.050\\
2454480.372487&  139.271&  0.050\\
2454705.645953&  139.049&  0.048\\
2454780.529216&  139.152&  0.052\\
2454819.336946&  139.207&  0.036\\
2455059.684154&  139.053&  0.031\\
2455149.486943&  139.131&  0.047\\
2455207.337039&  139.026&  0.034\\
2455232.321077&  139.059&  0.036\\
2455415.734837&  139.058&  0.032\\
2455439.622078&  138.995&  0.027\\
2455485.577689&  138.927&  0.032\\
2455503.470979&  138.867&  0.139\\
2455776.741087&  138.965&  0.034\\
2455821.588602&  138.997&  0.025\\
2455882.542281&  139.101&  0.047\\
2456139.725335&  138.991&  0.040\\
2456276.485443&  139.163&  0.039\\
2456519.736731&  138.956&  0.024\\
2456574.672228&  139.197&  0.047\\
2456956.570319&  139.252&  0.078\\
\hline
\end{tabular}           
\end{table}            

\begin{table}[ht]                   
\caption{HE~0432$-$0923}         
\centering                       
\begin{tabular}{lcr}
\hline\hline
HJD & RV & RV$_{err}$  \\
    & (km~s$^{-1}$) & (km~s$^{-1}$)\\
\hline
2454338.702068&  $-$64.948&  0.983\\
2454373.722874&  $-$64.876&  1.097\\
2454396.715153&  $-$66.294&  1.602\\
2454406.637662&  $-$64.942&  1.710\\
2454459.590478&  $-$62.264&  2.232\\
2454480.431485&  $-$65.026&  1.202\\
2454516.372208&  $-$64.832&  0.592\\
2454780.591463&  $-$64.462&  1.896\\
2454819.450079&  $-$64.460&  1.351\\
2455075.713863&  $-$65.352&  1.544\\
2455176.552628&  $-$64.543&  0.684\\
2455232.419859&  $-$65.290&  1.106\\
2455531.609740&  $-$64.654&  1.303\\
2455620.393797&  $-$63.236&  2.281\\
2455944.449119&  $-$66.678&  2.308\\
2456191.699846&  $-$64.856&  1.104\\
2456545.689792&  $-$64.892&  1.340\\
\hline
\end{tabular}           
\end{table}

\begin{table}[ht]                   
\caption{HE~0442$-$1234}         
\centering                       
\begin{tabular}{lcr}
\hline\hline
HJD & RV & RV$_{err}$  \\
    & (km~s$^{-1}$) & (km~s$^{-1}$)\\
\hline
2454338.746568&  231.895&  0.043\\
2454373.755033&  231.272&  0.022\\
2454396.675349&  231.387&  0.045\\
2454406.673221&  231.488&  0.047\\
2454480.469787&  231.625&  0.037\\
2454496.400729&  231.149&  0.135\\
2454516.413321&  231.777&  0.020\\
2454705.704003&  230.516&  0.030\\
2454780.549248&  230.670&  0.046\\
2454793.526401&  230.398&  0.173\\
2454819.485376&  231.059&  0.041\\
2454852.550506&  231.299&  0.045\\
2455059.709842&  254.627&  0.039\\
2455079.721105&  255.868&  0.030\\
2455126.674591&  252.221&  0.032\\
2455149.614597&  250.514&  0.037\\
2455171.540400&  248.975&  0.021\\
2455207.416924&  246.988&  0.029\\
2455232.382885&  245.710&  0.030\\
2455439.745696&  240.146&  0.032\\
2455478.744628&  239.853&  0.067\\
2455531.646433&  239.862&  0.033\\
2455821.705466&  236.600&  0.024\\
2455944.486938&  236.352&  0.046\\
2455971.360048&  236.071&  0.040\\
2456191.737737&  234.847&  0.023\\
2456574.706961&  233.444&  0.020\\
2456956.723997&  231.920&  0.055\\
\hline
\end{tabular}           
\end{table}            

                        
\begin{table}[ht]                   
\caption{HE~0524$-$2055}         
\centering                       
\begin{tabular}{lcr}
\hline\hline
HJD & RV & RV$_{err}$  \\
    & (km~s$^{-1}$) & (km~s$^{-1}$)\\
\hline
2454396.640800&  255.070&  0.185\\
2454406.717056&  255.306&  0.110\\
2454480.496860&  255.712&  0.098\\
2454820.481773&  255.614&  0.142\\
2455149.688274&  255.624&  0.084\\
2455207.491478&  255.546&  0.104\\
2455503.608363&  255.614&  0.551\\
2455531.544536&  255.478&  0.171\\
2455882.598085&  255.358&  0.108\\
2456005.376504&  255.236&  0.360\\
2456207.754219&  255.340&  0.082\\
2456603.661353&  255.443&  0.088\\
2456956.750015&  255.185&  0.196\\
\hline                          
\end{tabular}           
\end{table}            

\begin{table}[ht]                   
\caption{HE~1044$-$2509}         
\centering                       
\begin{tabular}{lcr}
\hline\hline
HJD & RV & RV$_{err}$  \\
    & (km~s$^{-1}$) & (km~s$^{-1}$)\\
\hline
2454909.489857&  342.628&  3.089\\
2454930.426948&  369.639&  1.671\\
2455174.767340&  379.567&  2.399\\
2455207.625989&  362.229&  1.939\\
2455620.563564&  384.868&  1.939\\
2455704.385256&  343.075&  2.077\\
2455712.405032&  335.836&  1.872\\
2455718.397062&  358.699&  2.243\\
2455722.397011&  375.517&  1.896\\
2455725.396840&  386.965&  3.807\\
2455730.395197&  385.629&  2.338\\
2455733.391004&  378.210&  3.132\\
2455738.401973&  354.412&  1.072\\
2456796.438819&  363.768&  1.346\\
\hline
\end{tabular}           
\end{table}

\begin{table}[ht]                   
\caption{HE~1105$+$0027}         
\centering                       
\begin{tabular}{lcr}
\hline\hline
HJD & RV & RV$_{err}$  \\
    & (km~s$^{-1}$) & (km~s$^{-1}$)\\
\hline
2454459.682587& 76.353&  1.799\\
2454464.668667& 76.746&  1.762\\
2454516.585694& 75.947&  1.514\\
2454909.650314& 75.405&  0.794\\
2455232.610390& 76.608&  0.991\\
2455344.429621& 76.620&  1.860\\
2455531.735132& 76.372&  1.831\\
2455662.477811& 76.280&  1.239\\
2456033.449115& 75.442&  1.403\\
\hline                        
\end{tabular}           
\end{table}            

\begin{table}[ht]                
\caption{HE~1127$-$1143}
\centering
\begin{tabular}{lcr}
\hline\hline
HJD & RV & RV$_{err}$  \\
    & (km~s$^{-1}$) & (km~s$^{-1}$)\\
\hline
2454459.748479&  229.676&  2.402\\
2454481.662602&  229.183&  1.172\\
2454964.473253&  228.850&  1.277\\
2455232.656932&  228.583&  1.005\\
2455620.602643&  229.334&  1.314\\
2455662.541359&  228.736&  1.308\\
2456458.454152&  229.734&  0.725\\
\hline                          
\end{tabular}                   
\end{table}                    
 
\begin{table}[ht]                  
\caption{HE~1219$-$0319}        
\centering
\begin{tabular}{lcr}
\hline\hline
HJD & RV & RV$_{err}$  \\
    & (km~s$^{-1}$) & (km~s$^{-1}$)\\
\hline
2454481.728699&  163.210&  3.750\\
2454625.460187&  162.919&  1.680\\
2455620.657226&  163.474&  2.343\\
2456090.413526&  161.166&  2.464\\
2456652.702430&  161.310&  2.342\\
\hline
\end{tabular}
\end{table}

\begin{table}[ht]
\caption{HE~1430$+$0053}
\centering
\begin{tabular}{lcr}
\hline\hline
HJD & RV & RV$_{err}$  \\
    & (km~s$^{-1}$) & (km~s$^{-1}$)\\
\hline
2454219.667720&  $-$107.225&  0.276\\
2454314.443374&  $-$108.199&  0.203\\
2454459.780220&  $-$107.456&  0.161\\
2454464.791534&  $-$107.738&  0.217\\
2454480.776288&  $-$107.848&  0.177\\
2454625.420967&  $-$107.371&  0.217\\
2454930.698058&  $-$107.699&  0.173\\
2454951.700884&  $-$107.748&  0.256\\
2454987.473997&  $-$107.662&  0.181\\
2455232.782505&  $-$107.709&  0.145\\
2455344.579755&  $-$107.138&  0.179\\
2455620.687317&  $-$107.719&  0.243\\
2455662.702207&  $-$107.560&  0.173\\
2455704.614991&  $-$107.359&  0.157\\
2455738.545590&  $-$108.383&  0.218\\
2455776.455326&  $-$108.773&  0.204\\
2456005.755704&  $-$108.120&  0.622\\
2456033.623529&  $-$108.201&  0.206\\
2456078.504562&  $-$107.171&  0.187\\
2456712.726996&  $-$107.900&  0.651\\
\hline                           
\end{tabular}                    
\end{table}                     


\begin{table}[ht]                   
\caption{HE~1523$-$0901}         
\centering                       
\begin{tabular}{lcr}
\hline\hline
HJD & RV & RV$_{err}$  \\
    & (km~s$^{-1}$) & (km~s$^{-1}$)\\
\hline
2454219.705449&  $-$163.546&  0.016\\
2454254.642226&  $-$163.593&  0.039\\
2454285.507316&  $-$163.348&  0.015\\
2454314.462351&  $-$163.385&  0.012\\
2454625.508366&  $-$163.091&  0.010\\
2454930.677427&  $-$163.244&  0.016\\
2454951.718880&  $-$162.971&  0.017\\
2454964.675531&  $-$162.951&  0.022\\
2454987.509060&  $-$163.041&  0.014\\
2455344.542474&  $-$162.849&  0.013\\
2455415.386096&  $-$163.549&  0.010\\
2455439.363821&  $-$163.724&  0.016\\
2455620.738975&  $-$162.875&  0.017\\
2455662.681656&  $-$163.110&  0.012\\
2455704.633870&  $-$163.372&  0.016\\
2455738.564258&  $-$163.723&  0.020\\
2455776.438140&  $-$163.732&  0.016\\
2456005.728789&  $-$163.530&  0.020\\
2456078.539561&  $-$163.442&  0.015\\
2456307.786409&  $-$163.457&  0.016\\
2456351.717323&  $-$163.347&  0.014\\
2456372.687769&  $-$163.477&  0.014\\
2456399.619896&  $-$163.371&  0.015\\
2456426.543882&  $-$163.298&  0.021\\
2456474.451216&  $-$163.108&  0.011\\
2456488.426825&  $-$162.960&  0.024\\
2456489.500440&  $-$163.000&  0.015\\
2456520.424407&  $-$162.818&  0.020\\
2456529.377789&  $-$162.938&  0.017\\
2456664.770980&  $-$163.461&  0.012\\
2456685.790500&  $-$163.626&  0.011\\
2456712.759468&  $-$163.414&  0.020\\
2456796.646910&  $-$162.880&  0.011\\
2456813.610868&  $-$162.988&  0.015\\
\hline                           
\end{tabular}                    
\end{table}                     

\begin{table}[ht]                   
\caption{CS~22892$-$052}         
\centering                       
\begin{tabular}{lcr}
\hline\hline
HJD & RV & RV$_{err}$  \\
    & (km~s$^{-1}$) & (km~s$^{-1}$)\\
\hline
2454314.576915&   13.178&  0.053\\
2454338.541838&   13.553&  0.136\\
2454373.414188&   13.559&  0.053\\
2454390.364640&   13.579&  0.079\\
2454625.684458&   13.594&  0.058\\
2454665.611641&   13.917&  0.063\\
2454705.568239&   13.747&  0.051\\
2455009.693164&   13.480&  0.098\\
2455059.511399&   13.695&  0.076\\
2455174.314885&   13.374&  0.050\\
2455415.522393&   13.471&  0.061\\
2455531.326532&   13.467&  0.108\\
2455704.706380&   13.742&  0.090\\
2455776.612557&   13.367&  0.049\\
2455822.600538&   13.505&  0.103\\
2455882.349889&   13.624&  0.051\\
2456139.644767&   13.413&  0.087\\
2456191.399426&   13.585&  0.099\\
2456488.673600&   13.584&  0.066\\
\hline                  
\end{tabular}           
\end{table}            

\begin{table}[ht]
\caption{HE~2224$+$0143}
\centering
\begin{tabular}{lcr}
\hline\hline
HJD & RV & RV$_{err}$  \\
    & (km~s$^{-1}$) & (km~s$^{-1}$)\\
\hline
2454314.626987&  $-$113.002&  0.070\\
2454338.455766&  $-$113.041&  0.061\\
2454373.470335&  $-$113.043&  0.113\\
2454625.646394&  $-$112.867&  0.147\\
2454665.630399&  $-$112.981&  0.118\\
2454705.522460&  $-$113.104&  0.115\\
2455009.711530&  $-$113.138&  0.099\\
2455059.529868&  $-$113.020&  0.056\\
2455070.548978&  $-$112.989&  0.159\\
2455175.444551&  $-$113.221&  0.079\\
2455344.685686&  $-$112.945&  0.119\\
2455415.497735&  $-$112.930&  0.102\\
2455439.426753&  $-$112.968&  0.103\\
2455503.356818&  $-$113.402&  0.352\\
2455531.410900&  $-$113.220&  0.136\\
2455704.677515&  $-$113.216&  0.095\\
2455776.507942&  $-$112.924&  0.093\\
2455796.580352&  $-$112.817&  0.094\\
2455882.328613&  $-$113.245&  0.106\\
2456078.720195&  $-$113.064&  0.081\\
2456140.610920&  $-$113.674&  0.186\\
2456191.379978&  $-$113.152&  0.059\\
2456488.693610&  $-$113.205&  0.084\\
2456956.553882&  $-$112.876&  0.303\\
\hline                           
\end{tabular}                    
\end{table}       

\begin{table}[ht]
\caption{HE~2244$-$1503}
\centering
\begin{tabular}{lcr}
\hline\hline
HJD & RV & RV$_{err}$  \\
    & (km~s$^{-1}$) & (km~s$^{-1}$)\\
\hline
2454338.580931&  147.980&  1.346\\
2454373.501898&  148.051&  1.007\\
2454390.393739&  148.275&  1.573\\
2454665.675576&  147.791&  1.192\\
2455059.563356&  147.724&  1.446\\
2455149.413670&  148.224&  1.470\\
2455415.573319&  148.194&  1.123\\
2455439.540189&  147.532&  1.306\\
2455478.531528&  147.567&  0.899\\
2455738.686456&  147.855&  0.872\\
2455776.556273&  147.947&  1.562\\
2455822.637317&  148.094&  1.569\\
2456191.476165&  148.111&  0.922\\
2456545.578750&  147.649&  1.362\\
\hline
\end{tabular}
\end{table}

\begin{table}[ht]
\caption{HD~221170}
\centering
\begin{tabular}{lcr}
\hline\hline
HJD & RV & RV$_{err}$  \\
    & (km~s$^{-1}$) & (km~s$^{-1}$)\\
\hline
2454314.636048&  $-$121.161&  0.010\\
2454338.524441&  $-$121.179&  0.009\\
2454373.528280&  $-$121.107&  0.009\\
2454390.345759&  $-$121.253&  0.010\\
2454406.586097&  $-$121.114&  0.011\\
2454480.397616&  $-$121.164&  0.013\\
2454625.656022&  $-$121.267&  0.012\\
2454705.500584&  $-$121.138&  0.017\\
2454780.452621&  $-$121.111&  0.012\\
2454793.471559&  $-$121.182&  0.011\\
2454820.348668&  $-$121.321&  0.014\\
2455009.719853&  $-$121.078&  0.011\\
2455059.695625&  $-$121.185&  0.009\\
2455071.639351&  $-$121.199&  0.017\\
2455171.399245&  $-$121.340&  0.008\\
2455344.693948&  $-$121.230&  0.011\\
2455415.506520&  $-$121.259&  0.012\\
2455439.446714&  $-$121.227&  0.008\\
2455503.346059&  $-$121.358&  0.014\\
2455531.470241&  $-$121.412&  0.011\\
2455704.717098&  $-$121.256&  0.009\\
2455738.710786&  $-$121.253&  0.018\\
2455776.728437&  $-$121.333&  0.012\\
2455796.566301&  $-$121.342&  0.011\\
2455859.532089&  $-$121.249&  0.009\\
2455892.424873&  $-$121.011&  0.029\\
2455915.320942&  $-$121.166&  0.077\\
2456140.740348&  $-$121.075&  0.010\\
2456241.340390&  $-$121.056&  0.013\\
2456488.721460&  $-$121.000&  0.010\\
\hline                   
\end{tabular}            
\end{table}             
                         
\begin{table}[ht]           
\caption{CS~30315$-$029} 
\centering
\begin{tabular}{lcr}
\hline\hline
HJD & RV & RV$_{err}$  \\
    & (km~s$^{-1}$) & (km~s$^{-1}$)\\
\hline
2454314.742671&  $-$168.937&  0.074\\
2454373.581743&  $-$169.408&  0.066\\
2454705.601838&  $-$169.668&  0.074\\
2455059.649134&  $-$169.963&  0.071\\
2455175.377253&  $-$168.867&  0.130\\
2455415.665073&  $-$169.830&  0.078\\
2455776.636347&  $-$169.582&  0.071\\
2455796.640879&  $-$169.184&  0.055\\
2455821.555408&  $-$169.165&  0.065\\
2455859.497621&  $-$169.214&  0.092\\
2455882.472080&  $-$169.087&  0.100\\
2456139.682491&  $-$169.008&  0.242\\
2456530.669987&  $-$169.196&  0.120\\
2456987.357305&  $-$169.738&  0.077\\
\hline
\end{tabular}
\end{table}

\clearpage

\section{Literature data for the programme stars}

\begin{table}[ht]
\begin{center}
\caption{Mean heliocentric radial velocities from the literature and total
  time-span covered for the single stars }
\label{tbl-B1}
\centering
\begin{tabular}{lrrrrl}
\hline\hline
Star ID & $\Delta$T Total&$\bar{RV}$ (this work)&$\bar{RV}$ (lit)& N & Ref \\
        & (days)          &(km~s$^{-1}$)           &(km~s$^{-1}$)  &   &  \\
\hline
\object{HD~20}         & 10011&$-$57.914 &$-$57.4 &3 & \citet{carney1986} \\
                       &      &          &$-$57.2 &13& \citet{carney2003} \\
                       &      &          &$-$57.5 &1 & \citet{barklem2005}\\
\hline
\object{CS~29497$-$004}& 4742 &$+$105.008&$+$105.1 &1 & \citet{barklem2005}\\
\hline
\object{CS~31082$-$001}& 5193 &$+$139.068&$+$139.1 & 8 & \citet{hill2002}   \\
                       &      &          &$+$138.9 &2 & \citet{aoki2003}\\
                       &      &          &$+$138.2&1 & \citet{tsangarides2003}\\
                       &      &          &$+$138.9 &1 & \citet{honda2004}      \\
                       &      &          &$+$139.4 &1 & \citet{barklem2005} \\
                       &      &          &$+$139.0 &1 & \citet{carrera2013}   \\
                       &      &          &$+$138.4 &1 & \citet{kordopatis2013} \\ 
                       &      &          &$+$138.9 &1 & \citet{roederer2014a}\\
\hline
\object{HE~0432$-$0923}& 3582 &$-$64.800 &$-$66.6 &1 & \citet{barklem2005} \\
\hline
\object{HE~0524$-$2055}& 4032 &$+$255.425&$+$255.3&1 & \citet{barklem2005} \\
\hline
\object{HE~1105$+$0027}& 3267 &$+$76.197 &$+$77.0 &1 & \citet{barklem2005} \\
\hline
\object{HE~1127$-$1143}& 3785 &$+$229.157&$+$228.5 &1 & \citet{barklem2005} \\
\hline
\object{HE~1219$-$0312}& 3885 &$+$162.416&$+$163.6 &1 & \citet{barklem2005}\\
                       &      &          &$+$163.1 &5 & \citet{hayek2009} \\
\hline
\object{HE~1430$+$0053}& 3942 &$-$107.749&$-$107.4 &1 & \citet{barklem2005} \\
\hline
\object{CS~22892$-$052}& 8788 &$+$13.549 &$+$13.1 &10& \citet{mcwilliam1995}\\ 
                       &      &          &$+$13.6 &1 & \citet{norris1996}   \\
                       &      &          &$+$12.5 &15&\citet{prestonsneden2001}\\
                       &      &          &$+$13.2 &1 & \citet{aoki2003}\\
                       &      &          &$+$12.7 &1 & \citet{honda2004} \\
                       &      &          &$+$14.5 &1&\citet{barklem2005}\\
                       &      &          &$+$13.3 &1 & \citet{bonifacio2009} \\
                       &      &          &$+$13.0 &2 & \citet{roederer2014a} \\
\hline
\object{HE~2224$+$0143}& 3968 &$-$113.085&$-$112.3&1 & \citet{barklem2005} \\
\hline
\object{HE~2244$-$1503}& 3960 &$+$147.928&$+$148.1&1 & \citet{barklem2005} \\
\hline
\object{HD~221170}     & 10921&$-$121.201&$-$119.0&4 & \citet{wilson1953}\\
                       &      &          &$-$123.7&3 & \citet{wallerstein1963}\\
                       &      &          &$-$121.8&18& \citet{carney2003} \\
                       &      &          &$-$120.7&1 &\citet{barklem2005}\\
                       &      &          &$-$121.7&1 & \citet{carney2008}\\
                       &      &          &$-$121.8&1 & \citet{soubiran2008}\\
\hline
\object{CS~30315$-$029}& 4741 &$-$169.346&$-$169.2 &1 & \citet{barklem2005} \\
\hline
\end{tabular}
\end{center}
\end{table}

\clearpage

\begin{table}[ht]
\caption{Mean heliocentric radial velocities for single stars not included in the programme}
\label{tbl-B2}
\centering
\begin{tabular}{lrrrl}
\hline\hline
Star ID & $\Delta$T Total&$\bar{RV}$  & N & Ref \\
        & (days)         &(km~s$^{-1}$)&   &  \\
\hline
\object{HD~115444} & 8812 &$-$27.6 & 3&\citet{griffin1982}\\
                   &      &$-$26.2 & 2&\citet{bartkevicius1992}\\
                   &      &$-$27.1 & 2&\citet{aoki2003}\\       
                   &      &$-$27.2 & 1&\citet{famaey2005}\\
\hline
\object{HD~122563} &21575 &$-$26.1 & 3&\citet{wilson1950}\\
                   &      &$-$26.5 & 6&\citet{wallerstein1963}\\
                   &      &$-$24.9 & 6&\citet{woolley1965}\\
                   &      &$-$26.0 & 1&\citet{bond1980}\\
                   &      &$-$26.0 & 1&\citet{roederer2008}\\
                   &      &$-$26.4 & 1&\citet{bonifacio2009}\\
                   &      &$-$25.6 & 2&\citet{hollek2011}\\
                   &      &$-$26.1 & 3&\citet{roederer2014a}\\

\hline
\end{tabular}
\end{table}

\end{appendix}

\end{document}